# The Physical State of Lunar Soil in the Permanently Shadowed Craters of the Moon

Jacob N. Gamsky[1], Philip T. Metzger[2]

[1] Student, Department of Physics and Astronomy, University of Kentucky, Lexington, KY 40506; email: jngams2@uky.edu
[2] Regolith Mechanics and Operations Lab, NASA, Kennedy Space Center, Florida, 32899; email: philip.t.metzger@nasa.gov

## ABSTRACT

The physical state of the lunar soil in the permanently shadowed craters of the moon is inferred from experimental investigation. The permanently shadowed craters do not undergo the same thermal cycling experienced by other parts of the moon and therefore could be slightly less compacted. This study is significant because excavating, roving, and landing interactions, along with the energy budgets and deployment schedules for associated technology, need to be scaled and designed properly. Results indicate that the degree of compaction due to thermal cycling is a function of the depth in the soil column.

## INTRODUCTION

The surface density of the regolith in the permanently shadowed craters of the moons poles is expected to vary with respect to other areas on the lunar surface. The surface density of lunar soil is predicted to significantly affect the erosion rate under a rocket exhaust plume due to the interaction of particle forces. Due to the low gravity and lack of atmosphere on the lunar surface jets of rocket exhaust (from landing spacecraft) spray the top layer of loose lunar regolith in all directions at approximately 2 km/s (Metzger, et al, 2009). There has been significant evidence of this occurring from data recovered from the deactivated Surveyor 3 spacecraft as a result of the Apollo 12 Lunar Module landings. The Apollo 12 LM landed 155 m away from the Surveyor spacecraft and pieces of the side facing the Apollo LM were returned to Earth. Data and analysis conducted on the pieces of the spacecraft showed extensive sandblasting, pitting, and cracking (Immer, et al, 2009). This poses a significant problem for future lunar missions. With this said, the permanently shadowed craters of the moon must be studied because future missions plan to land in or around these craters. It is vital to know the density of the regolith in these regions so excavating, roving, and landing interactions, along with the energy budgets and employment schedules for related technology, can be scaled and calculated properly. This technology advancement is already in progress and cannot wait for a robotic lander to make in-situ measurements.

The surface density of the regolith in these craters may be significantly less compacted when compared with unshadowed areas of the moon. The theory behind that statement deals with the thermal cycling of granular materials, which is known to be a very efficient compactor of soil (Chen, et al, 2006). The temperature of the lunar days (which last for 2 weeks) can reach around 123˚C and drop as low as -233˚C during the lunar night (Heiken, 1991). These extreme temperature swings compact the soil because particle expansion in the hot period results in high tangential stress that drives particle reorientation, and particle contraction in the cold period relaxes the soil under gravity, permanently consolidating the grains (Chen, et al, 2006). The thousands of thermal cycles that have occurred throughout history have compacted the lunar regolith into a very dense state. The angle of the sun on these permanently shadowed craters prohibits any light (and thus heat) from entering. As a result, the lunar regolith particles in the craters do not undergo the extreme temperature cycles that is experienced by the regolith in the other areas of the moon. The second process that compacts lunar regolith is vibration. The research conducted tested which compaction process was the dominant process on the lunar surface.

## EXPERIMENTAL METHODS

Experimental methods and equipment were developed to test the two dominant regolith compaction processes on the moon (thermal cycling and vibration) using the lunar regolith simulant JSC-1A. Three groups of ten cylindrical beakers with un-compacted JSC-1A (500 ml of simulant in each beaker) were heated (using a furnace), vibrated (using a vibration table), and left undisturbed in order to simulate the lunar compaction processes. In addition; each beaker had an overlying weight to represent the overburden weight of the soil at different depths (down to 1 meter, Figure 1). These mass values were calculated using the density equation found from the Apollo missions (Heiken, 1991),

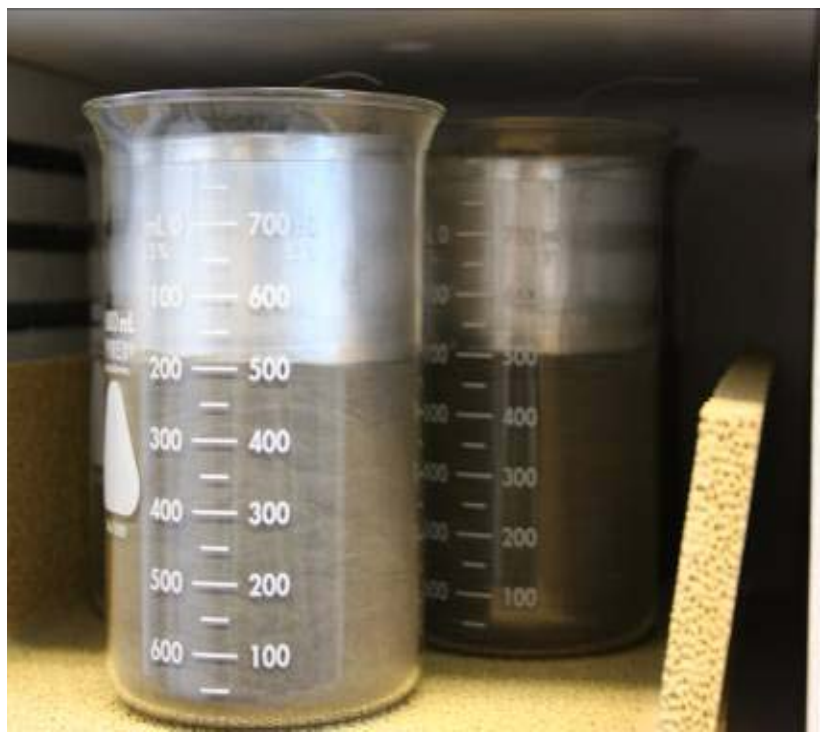

**Figure 1. 500 ml of JSC -1A with aluminum container filled with weight to represent depth below surface.**

$$\rho = 1.92\frac{z+12.2}{z+18} \qquad (1).$$

Where ρ is the density and z is the depth from the surface. This data is shown in Table 1.

One set of 10 beakers was thermal cycled; a second set of 10 beakers was vibrated; a third set was left undisturbed as a control group. After each thermal cycle and vibration time interval the compaction was measured using a laser distance measuring device mounted into a fixed position above the beakers. Testing of the vibrational group ended when all samples reached steady state. Due to the long times required for thermal cycling (and schedule limitations), testing of the thermal cycled group was ended after five cycles even though they had not yet reached steady state. This was sufficient to see the trends, but more testing is needed to reach steady state. It should be noted that vibrations do not match the real lunar environment, but the behavior of the soil was studied under a variety of the vibrational amplitudes to gain insight into the more complex, realistic cases.

## EXPERIMENTAL RESULTS

The vibration tests returned several pieces of useful data. The data was analyzed in a variety of configurations to give an accurate representation of the overall compaction process and to allow for each variable to be analyzed. The key results returned from the vibration studies were that the relative density of the regolith increased as the vibration frequency increased, but the compaction was not necessarily a function of depth.

**Table 1. The depths tested and overburden weight calculated at each depth.**

| Beaker | Simulated Lunar Depth (m) | Overburden Mass (kg) |
|---|---|---|
| 1 | 0 | 0 |
| 2 | 0.1 | 0.1778 |
| 3 | 0.2 | 0.3746 |
| 4 | 0.3 | 0.5814 |
| 5 | 0.4 | 0.7943 |
| 6 | 0.5 | 1.0114 |
| 7 | 0.6 | 1.2315 |
| 8 | 0.7 | 1.4539 |
| 9 | 0.8 | 1.6782 |
| 10 | 0.9 | 1.9039 |

The data showed that each beaker compacted monotonically to approximately the same levels as one another, determined by the vibrational acceleration, but independent of over burden weight (simulated depth), as illustrated in Figure 2. Figure 2 shows the graph of relative density of the samples for specific overburden weights (depths) for each vibration level (g's). This is a comparison graph to contrast the compaction at each overburden weight. Nearly every sample of JSC-1A increased in density at each vibration and nearly every sample compacted to approximately the same relative density.

The two abnormalities in this data plot occurred in beakers 1 (0 kg overburden weight) and 6 (1.01 kg overburden weight). The simulant that contained no overburden weight began convecting at the 1.41 $m/s^2$ acceleration value. This explains why the relative density started to decrease during that test. This experimental finding agrees with the well known observations of the Apollo astronauts who noted that the top few centimeters (no overburden weight) of lunar regolith was very loose, but became very compacted as the depth into the surface increased.

In addition, the abnormality in the beaker that contained 1.01 kg of overburden weight is not fully understood, however the belief is that the mass of the overburden weight was in resonance with the vibrational table during certain acceleration values and this may explain the usual behavior exhibited by this sample. As illustrated in Figure 2, the relative density of this sample increased significantly (when compared to the other samples) during the acceleration values of 1.09 $m/s^2$ and 1.41 $m/s^2$. Also, the relative density decreased in this sample during the two highest acceleration value tests (2.05 $m/s^2$ and 2.25 $m/s^2$ ). It should be noted that this was a repeatable condition, not an experimental error.

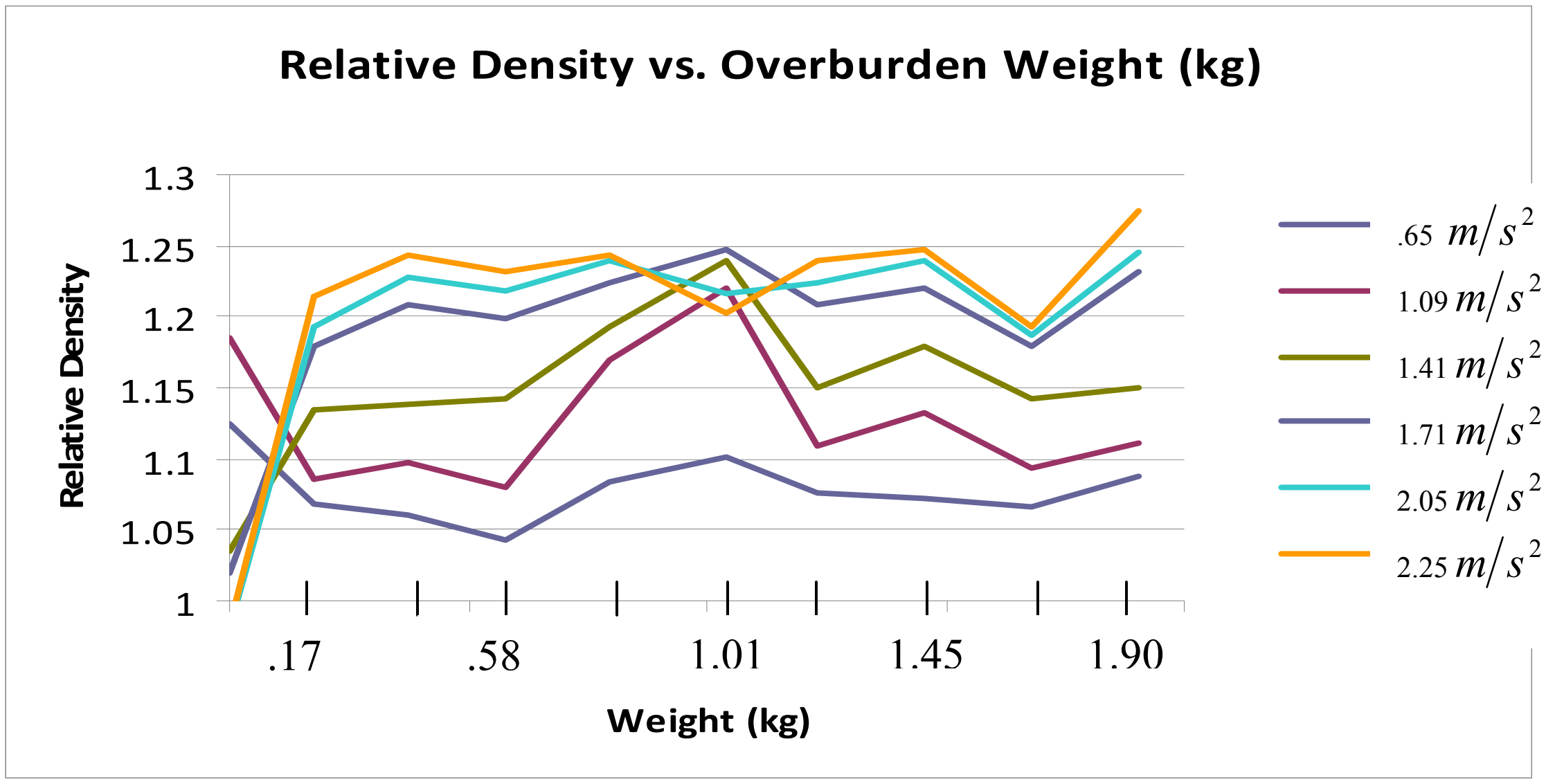

**Figure 2. Steady State relative density vs. overburden weight at each acceleration value.**

The most significant result was in regards to the thermal cycled beakers of JSC-1A lunar simulant. The 10 beakers with respective overlying weight went through 5 (time intensive) thermal cycles (between ambient and 250˚C). The results shown in table 2 give evidence to the fact that thermal cycling compacts (distance from starting height) JSC-1A and it is a function of depth. At every subsequent depth the compaction is greater (except for the .4 -.5 beakers) then the previous depth. This was a previously untested result and the *first* study performed into the compaction of a lunar simulant as a function of depth.

**Table 2. Compaction of simulant after 5 thermal cycles.**

| Simulated Lunar Depth (m) | Compaction (mm) |
|---|---|
| 0 | 0.263 |
| 0.1 | 0.3125 |
| 0.2 | 0.8125 |
| 0.3 | 0.875 |
| 0.4 | 1.355 |
| 0.5 | 1.325 |
| 0.6 | 1.0625 |
| 0.7 | 1.3125 |
| 0.8 | 1.3625 |
| 0.9 | 1.3812 |

Figure 3 shows a comparison of the experimental results from thermal cycling the regolith samples vs. the predicted density as a function of depth calculated from the equation found in the lunar sourcebook (Heiken, 1991). This representation axes are skewed to show the comparison.

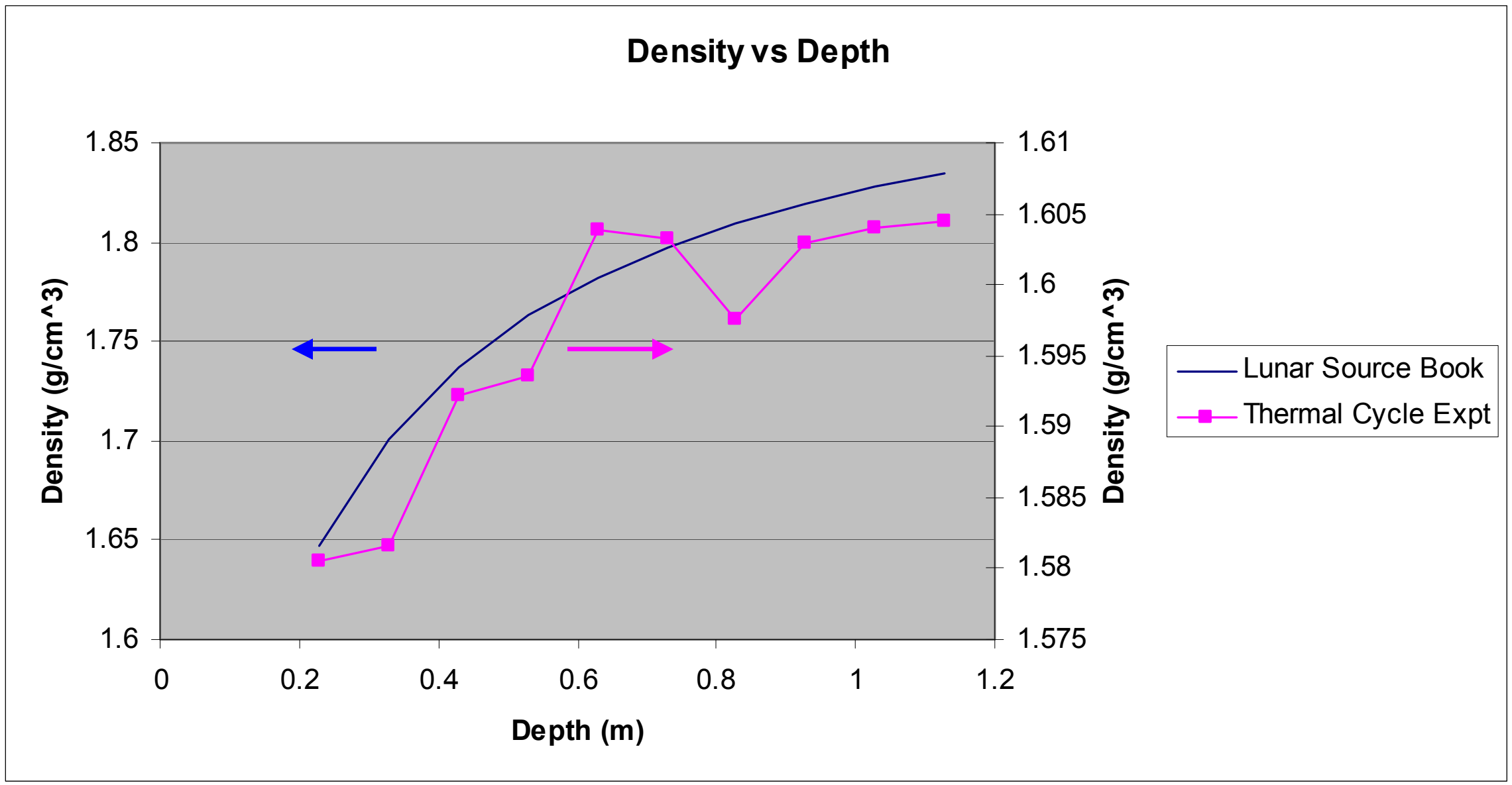

**Figure 3.** A plot of actual vs. experimental values (as a result of thermal cycling) for density as a function of depth on the lunar surface.

## CONCLUSIONS

Research into the compaction processes of lunar regolith is vital for future missions to the permanently shadowed craters and surrounding areas of the moon. In order to build a permanent outpost, engineers have to design equipment, methods, and schedules for building, excavating, landing, roving, and numerous other processes needed for survival on the lunar surface. The density of the soil in these different areas is a key piece of information needed for their designs.

This research has given evidence to the fact that both vibration and thermal cycling are effective at compacting lunar regolith, however, previous studies have only investigated vibrational effects. It has been shown that vibrating samples of lunar regolith with overburden weight decompacts the very top layer of the regolith (0 kg of overburden weight), but increases the relative density of the subsequent layers. The vibrational tests increased the relative density of the samples more-or-less uniformly regardless of depth. Consequently, Eq. 1 does not appear to be a valid representation of the lunar soil, unless it can be explained through thermal cycling.

In addition, thermal cycling does increase the relative density of lunar regolith as a function of depth (this agrees with Eq. 1 to as far down as temperature waves reach). However, more thermal cycles are needed to confirm this result.

The thermal wave is known to reach only 1 m into the soil, but since all layers of soil were at one time in the top layer, all have been compacted by thermal cycling. The combination of realistic vibration and thermal cycling environments over appropriate time scales is needed to explain the density of soil in each layer. Since densification is a time-dependent process (both vibration and thermal cycling having their own time scales), the densities in the lunar strata layer may serve as a useful dating method to understand how long each layer existed at the surface before being covered over and thus may help unravel the stratigraphy and record of impacts.

Since it has been shown that thermal cycling effectively compacts the lunar soil, and since all compaction is monotonic, we infer that the soil in the permanently shadowed craters must be less compacted than elsewhere on the Moon. Further research is needed to quantify this statement. In addition, because the soil in the shadowed craters is less compacted, it is more porous and this may affect our estimates of its ice inventory. It also implies the soil will be weaker to excavate and more prone to erosion and cratering beneath a landing rocket (unless cemented with ice). Again, more research is needed to quantify these conclusions.